\documentclass[final,3p,times,twocolumn]{elsarticle}

\usepackage{amssymb}

\usepackage{amsmath}
\usepackage{multirow}
\usepackage{dcolumn}
\newcolumntype{.}{D{.}{.}{-1}}

\usepackage{color}

\usepackage{upgreek}

\journal{Sensors and Actuators A: Physical}

\begin{document}

\begin{frontmatter}



\title{Design of a CubeSat Payload to Test a Magnetic Measurement System for Space-borne Gravitational Wave Detectors}

\author[uca]{I. Mateos}	
\ead{ignacio.mateos@uca.es}
\author[ieec]{R. S\'anchez-M\'inguez}
\author[upc,ieec]{J. Ramos-Castro}


\address[uca]{Escuela Superior de Ingenier\'ia, Universidad de C\'adiz, 11519 C\'adiz, Spain}
\address[ieec]{Institut d'Estudis Espacials de Catalunya (IEEC), 08034 Barcelona, Spain}
\address[upc]{Departament d'Enginyeria Electr\`onica, Universitat Polit\`ecnica de Catalunya, 08034 Barcelona, Spain}

\begin{abstract}
Space observatories for gravitational radiation such as LISA are equipped with dedicated on-board  instrumentation capable of measuring magnetic fields with low-noise conditions at millihertz frequencies. The reason is that the core scientific payload can only operate successfully if the magnetic environment meets certain strict low-frequency requirements. With this purpose, a simplified version of the proposed magnetic measurement system for LISA has been developed for a six-unit CubeSat, which will make it possible to improve the technology readiness level (TRL) of the instrument. The special feature of the experiment is that the magnetic sensors integrated in the payload are magnetically shielded to low-frequency fluctuations by using a small cylindrical permalloy enclosure. This will allow the in-flight noise characterization of the system under the CubeSat orbit environment. Therefore, a CubeSat platform will offer the opportunity to measure the capability of the instrument and will guide the progress towards the improved magnetic measurement system for LISA. This article describes the principal characteristics and implementation of the CubeSat payload.

\end{abstract}


\end{frontmatter}


\section{Introduction}\label{intro}

The first direct detection of gravitational waves by ground-based laser interferometric observatories in 2015 opened a new era of gravitational astronomy along the high-frequency regime, i.e. over $1\,{\rm Hz}$\,\cite{bib:FirstGW, bib:Detectors}. In addition, the future space-based gravitational wave detector called LISA will complement the knowledge of the exotic sub-hertz gravitational Universe\,\cite{bib:LISAProposal},  which is not accessible from laser interferometric terrestrial observatories due to the effect of local gravitational disturbances and arm length limitations on instrument sensitivity. 

LISA has already been selected as the third large-class mission (L3) for the ESA Cosmic Vision science programme in June 2017. Thence, the next step is to advance the technical maturity of critical technology elements that will be part of  the spacecraft. The capabilities of some of these instruments have been successfully proved in its technology demonstrator, LISA Pathfinder\,\cite{bib:FirstLPFResults,bib:FinalLPFResults}. However, optimization of specific items is desirable due to the more demanding requirements in LISA. Ideally, the ongoing technology developments need to be at technology readiness level (TRL) of 6 or higher by the early 2020s, this is before the mission adoption review of the ESA science program. The reader can find more details about the current TRLs of the primary items of the mission in \,\cite{bib:LISAProposal}.

Due to the extremely tiny effect of a passing gravitational wave, the free-falling reference test mass, which is the core of the gravitational reference sensor (GRS), needs to be shielded from non-gravitational forces that can conceal the gravitational signal. Therefore, environmental conditions, which can behave as force noise sources, have to be monitored by several sensors, including magnetometers. This on-board magnetic instrumentation will help to effectively disentangle during post-processing the spurious  force on the test mass as a consequence of the surrounding field coupled with the non-zero magnetization and susceptibility of the test mass.



The current magnetic measurement system proposed for LISA is obviously based on LISA Pathfinder's, where mature fluxgate technology was selected due to its proven low-noise along the measurement bandwidth and its actual availability for space applications\,\cite{LPFMag,bib:DMU,bib:DDS_LTP}. However, as far as LISA is concerned, a number of further improvements need to be considered, which have derived in the study of alternative technologies to fluxgate magnetometers with the purpose of inferring a more  faithful  reconstruction  of  the  magnetic  field  and magnetic field gradient at  the  spacecraft. Some advantages brought on by using chip-scale sensing technology instead of bulky fluxgate magnetometers are briefly described: (i) a larger set of smaller sensors makes it possible to better estimate the noise projection from the magnetic field sources; (ii) magnetometers with sufficiently low magnetic impact on the spacecraft environment, i.e., tiny quantities of ferromagnetic materials, allow to  mount them  closer to the free-falling test mass;  (iii) the deviations from the foreseen position of the sensor head, known as spatial uncertainty, are also reduced with the use of a more compact magnetometer. It must be remarked that the trade-off between smaller and yet sensitive magnetometers is critical since noise curves steeply rise for tinier magnetometers and towards lower frequencies\,\cite{Ripka20038}. Besides, the measurement bandwidth for LISA is extended down to 0.1 mHz, which increases the complexity of reaching the noise level at a frequency one order of magnitude lower than that for LISA Pathfinder. At present, two sensing techniques have been studied for the ongoing magnetic measurement system for LISA, on-chip anisotropic magnetoresistances (AMRs)\,\cite{bib:AMRpaper,bib:Interpolation} and novel atomic magnetometers\,\cite{bib:atomicLISA}.

This paper describes the development of a low-power small-sized model based on the currently proposed magnetic measurement system for LISA using AMRs and according to the CubeSat design specification.  In principle, the purpose is to serve the $^3$Cat-2 six-unit CubeSat\,\cite{bib:3cat-2}, coordinated by the Remote Sensing Laboratory at the Universitat Polit\`ecnica de Catalunya  (UPC), as a platform for in-orbit demonstration of the instrument. Nevertheless, the payload can be easily integrated in other CubeSat. This will make the validation of the novel technological concept under operational mission conditions possible. 



\section{CubeSat requirements for the payload}\label{sec:req}

Low-cost satellites have especially demanding constraints regarding power consumption, physical dimensions, power lines, weight, and on-board data handling, which can restrict the performance of the on-board instrumentation.  Table\,\ref{tab:CubeSatReq} shows the principal requirements imposed by the $^3$Cat-2 project for our electronic board. Thus, the initial design\,\cite{bib:AMRpaper} was slightly changed by trading off magnetometer characteristics, such as range or bandwidth, for CubeSat requirements.

\begin{table}[ht!]
\caption{Specific CubeSat requirements for the payload.}
\centering
\begin{tabular}{l  c}
\hline \hline
{\bf Parameter} & {\bf Value} \\ 
	\hline 
	Power line & +5V \\ 
	Power consumption & $< 0.5\,{\rm W}$ \\
	Physical dimensions & 95.89 x 90.17 x  30 mm$^3$\\ 
	Mass & $\leq 100\,{\rm g}$ \\
	Temperature range & $-40$ to $85\,^{\circ}{\rm C}$  \\ 
	Max. cont. operat. time & 3 hours\\ 
	Bus interface & I$^2$C \\ 
	\hline
	\hline
\end{tabular}
\label{tab:CubeSatReq}

\end{table}

Moreover, the payload is required to work in autonomous operations. Consequently, a microcontroller ($\upmu$C) and EEPROM devices  were added to the front-end electronics in order to store the data until requested from the CubeSat's on-board computer (OBC).

\section{Specifications of the CubeSat payload}

\begin{figure}[ht!]
\centering
   \includegraphics[width=0.5\textwidth]{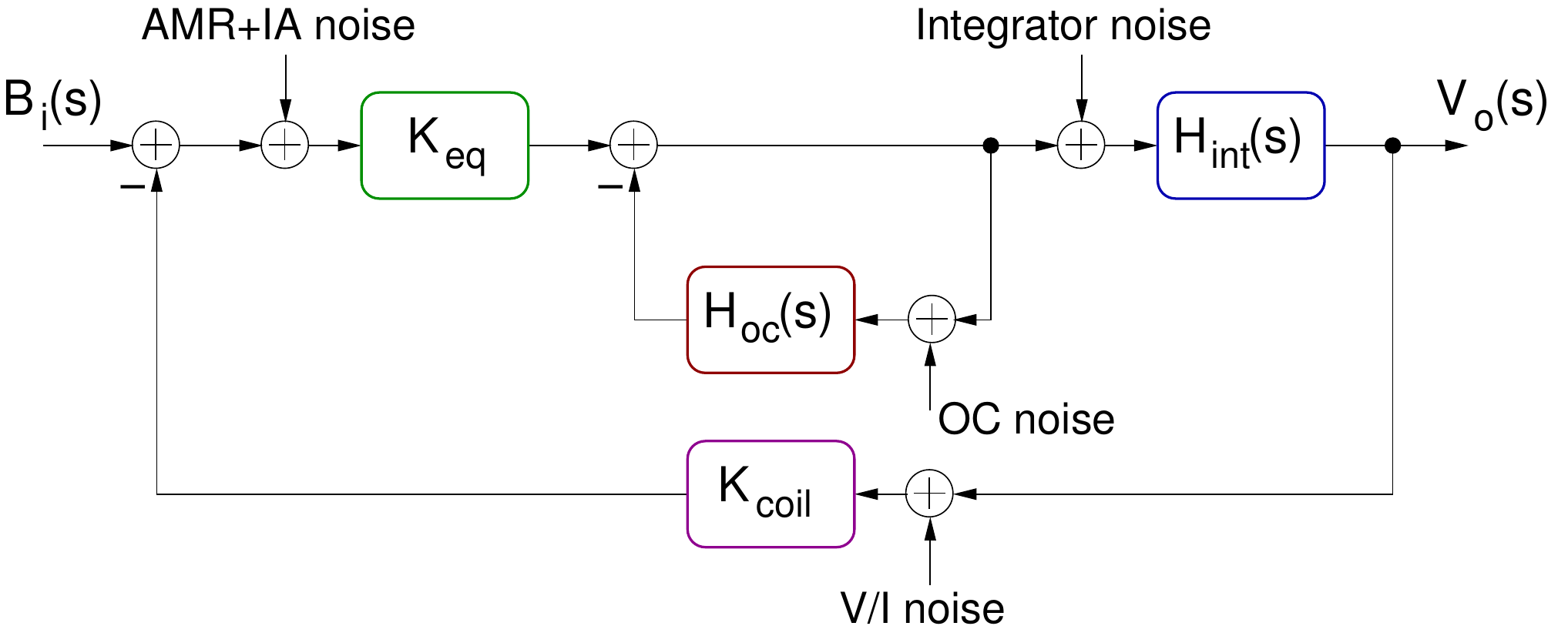}
   \caption{\label{fig:TF_circuit} Scheme of the analog signal processing including noise sources and feedback loops.}
\end{figure}

\begin{figure*}[bt]
\centering
   \includegraphics[width=1\textwidth]{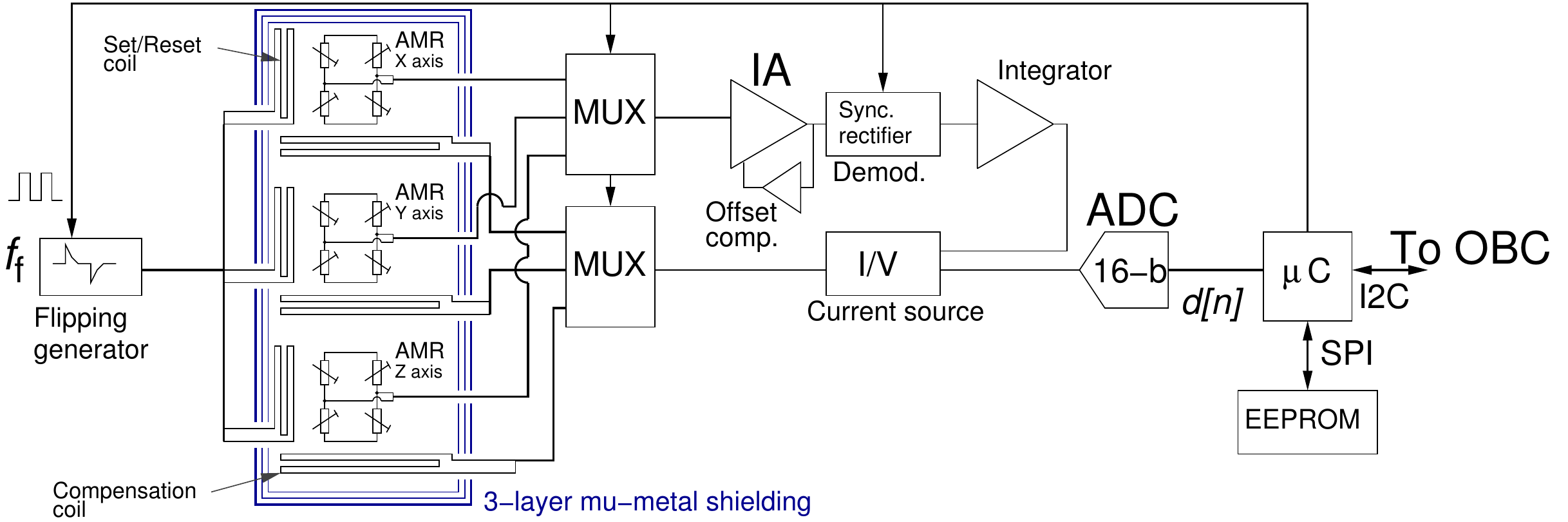}
   \caption{\label{fig:BlockDiagram} Block diagram of the CubeSat payload. A mu-metal shielding encloses the AMR sensors for in-orbit magnetic noise measurements of the system.}
\end{figure*}

The sensing element is composed of a Wheatstone bridge of AMRs, whose analog signal is modulated, amplified by a precision instrumentation amplifier with a gain of $100$, synchronously demodulated using  a phase sensitive detector, integrated, and quantized by an analog-to-digital converter (ADC). The noise estimate of each stage was quantified analytically and verified by on-ground measurements. Among the different noise contributions of the circuit, the one from the instrumentation amplifier becomes relevant at frequencies above 1 Hz, but the critical source in the bandwidth of interest is clearly the sensor's intrinsic $1/f$ noise\,\cite{bib:AMRpaper}. The dominant flicker noise of AMRs together with their significant thermal dependences come out as excess noise in the millihertz band. Thus, dedicated modulation and closed-loop techniques are performed in order to ameliorate the noise curve of the system along the LISA bandwidth\,\cite{bib:AMRpaper, bib:flipping, bib:AMRtemp}. The scheme of the analog conditioning circuit considering the most relevant noise sources is given in Figure\,\ref{fig:TF_circuit}, and the closed-loop response of the system is

\begin{align}
\nonumber
V_{\rm o}(s) &= \Bigl[K_{\rm eq}\Bigl(B_{\rm i}(s) + n_{\rm AMR+IA} - n_{\rm V/I}K_{\rm coil}\Bigr) \\ \nonumber
& + n_{\rm int}\Bigl(1 + H_{\rm oc}(s)\Bigr) - n_{\rm oc}H_{\rm oc}(s)\Bigr]\\
&\times \frac{  H_{\rm int}(s)}{1 + H_{\rm oc}(s) + K_{\rm eq}   K_{\rm coil} H_{\rm int}(s)},
\end{align}

\noindent where $B_{\rm i}(s)$ is the magnetic field to be measured, $n_{\rm AMR+IA}$, $n_{\rm V/I}$, $n_{\rm int}$, and $n_{\rm oc}$ are the input noise of each stage (AMR, instrumentation amplifier, voltage-to-current converter, integrator, and offset compensation), $K_{\rm eq}$ is the product of the AMR bridge sensitivity and the instrumentation amplifier gain, $K_{\rm coil}$ is the gain of the voltage-to-current converter multiplied by the compensation coil ratio , and $H_{\rm int}(s)$ and $H_{\rm oc}(s)$ are the transfer functions of the integrator and offset compensation. The block diagram of the payload is shown in Figure\,\ref{fig:BlockDiagram}.

 Due to the aforementioned requirements in Section\,\ref{sec:req}, the most relevant modifications from the original design that affect the performance of the magnetic measurement system are: (i) use of operational amplifiers with lower power consumption, albeit higher noise (OP470 instead of the OP27 used in the original design);  (ii) magnetic field range reduction (lower feedback current); and (iii) the three axes of the magnetometer are not powered simultaneously, thus, only data from one axis can be acquired in each run. Besides, the system can not be characterized at the lower end of the LISA bandwidth ($0.1\,{\rm mHz}$) with the mission operation constraint for the CubeSat of three hours of data per run. Consequently, we have to settle for this limitation estimating the noise spectral densities with around ten points averaging at 1\,mHz. It is important to note that in spite of these adjustments, the in-flight noise experiment will allow to observe the behavior of the 1/$f$ noise corner frequency, which is the figure of merit together with the noise floor of the system. A brief overview of the principal characteristics of the CubeSat payload is offered below. 

\subsection{Magnetic shielding}

The AMR magnetometers were placed inside a small cylindrical enclosure for the purpose of shielding the environmental magnetic field down to the nanotesla range, thus allowing for in-flight low-frequency noise characterization of the magnetometers. The local spacecraft field will be dominated by the Earth geomagnetic field and the local spacecraft magnetic sources,\footnote{The Earth magnetic field at a LEO altitude of 400\,km varies between $25\,\rm {\upmu T}$ and $50\,\rm {\upmu T}$\,\cite{bib:LEOfield}.} such as the the magnetic field generated by the active magnetorquer used for attitude control. Nevertheless, the magnetorquer can be turned off during the noise characterization experiment if necessary.

\begin{figure}[ht!]
  \centering
 \includegraphics[width=1.1\linewidth]{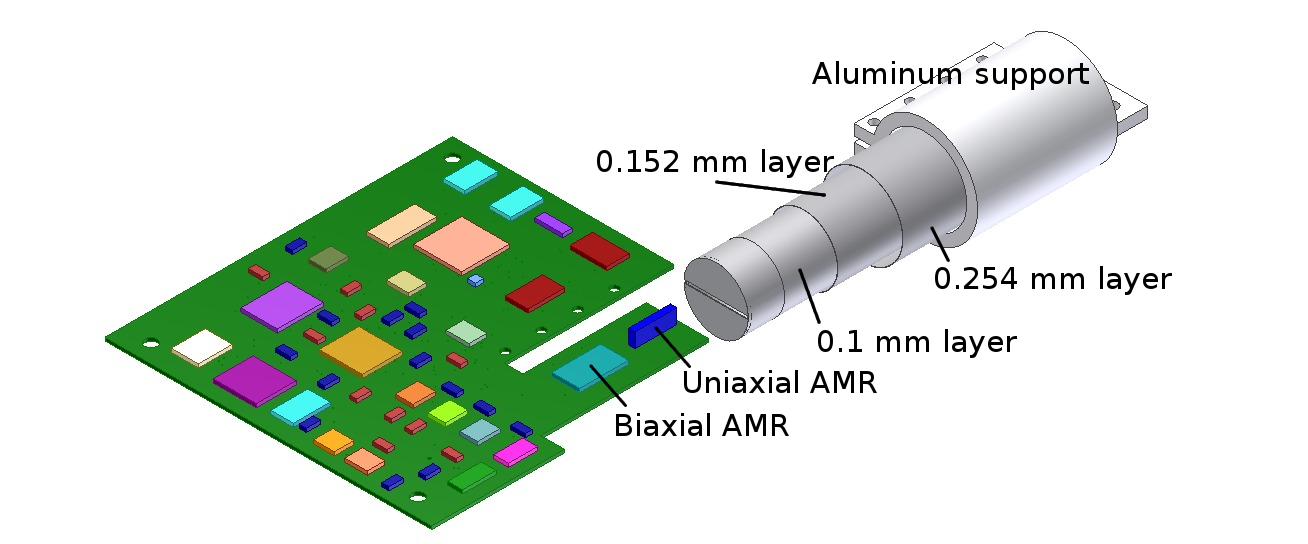}
 \vspace{0.6cm}
 \includegraphics[width=0.47\textwidth]{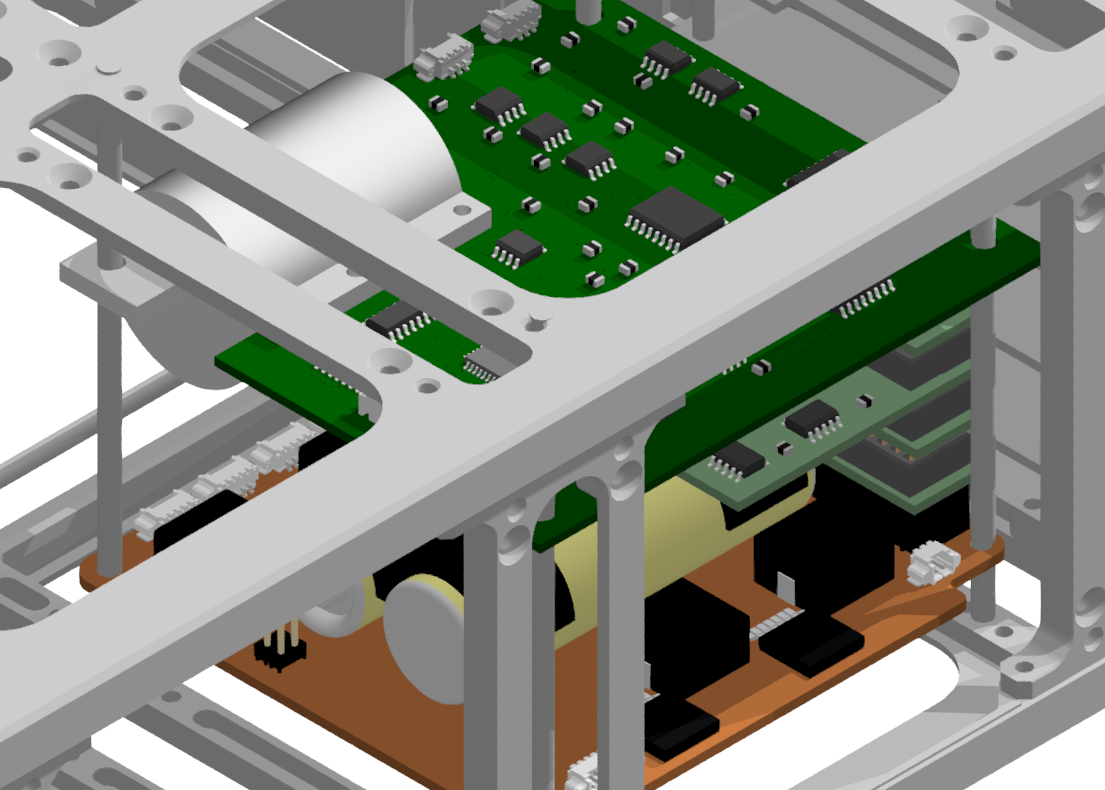}
 \caption {Top: CubeSat payload with the three mu-metal layers for the magnetic shielding. Bottom: Drawing of the magnetic measurement system with the mu-metal shield integrated in one standard slot of the CubeSat. For the sake of clarity, only the electronics boards for one unit (1U) of the 6U CubeSat structure (ISISpace\,\cite{bib:ISIS}) is shown. The  CubeSat power module (GomSpace/NanoPower P31U\,\cite{bib:GomSpace}) is allocated right under our payload.}
 \label{fig:PayloadCubeSat}
\end{figure}

The magnetic shield consists of three nested mu-metal layers surrounding the sensor head of the magnetometer. The outermost cylinder is a 0.254 mm-thick layer, the centered one has a thickness 0.152 mm, and the innermost shield is a 0.1\,mm-thick layer. These three layers are separated by gaps of around 4.5 mm each and assembled to the board  by means of a cylindrical aluminum structure that has a diameter of 29\,mm and is 38.6-mm-long (see Figure\,\ref{fig:PayloadCubeSat}). 


\subsection{Low-frequency noise performance}


As measured in-flight by the LISA Pathfinder magnetometers, low-frequency magnetic field fluctuations across the LISA spacecraft are expected to arise from the interplanetary magnetic field with an amplitude spectral density of $\simeq 100\,{\rm nT\,Hz^{-1/2}}$ at 0.1 mHz\,\cite{bib:FirstLPFResults,bib:Bip1}. Whereas, the magnetic field gradient fluctuations are predicted to be dominated by the spacecraft's magnetic sources\,\cite{bib:TMproperties}. Therefore, since the LISA requirements at subsystem level and the magnetic sources' distribution in the spacecraft are not yet entirely defined, as a first approach, the noise performance of the magnetic measurement system must be at least an order of magnitude less noisy than the expected interplanetary magnetic spectral density to be detected at the low-frequency end of the band. This implies a stability requirement in the measurement system of  $S^{1/2}_{B,\rm system}\leq 10\,{\rm nT\,Hz^{-1/2}}$ at $0.1\,{\rm mHz}$.

\begin{figure}[ht!]
  \centering
 \includegraphics[width=0.48\textwidth]{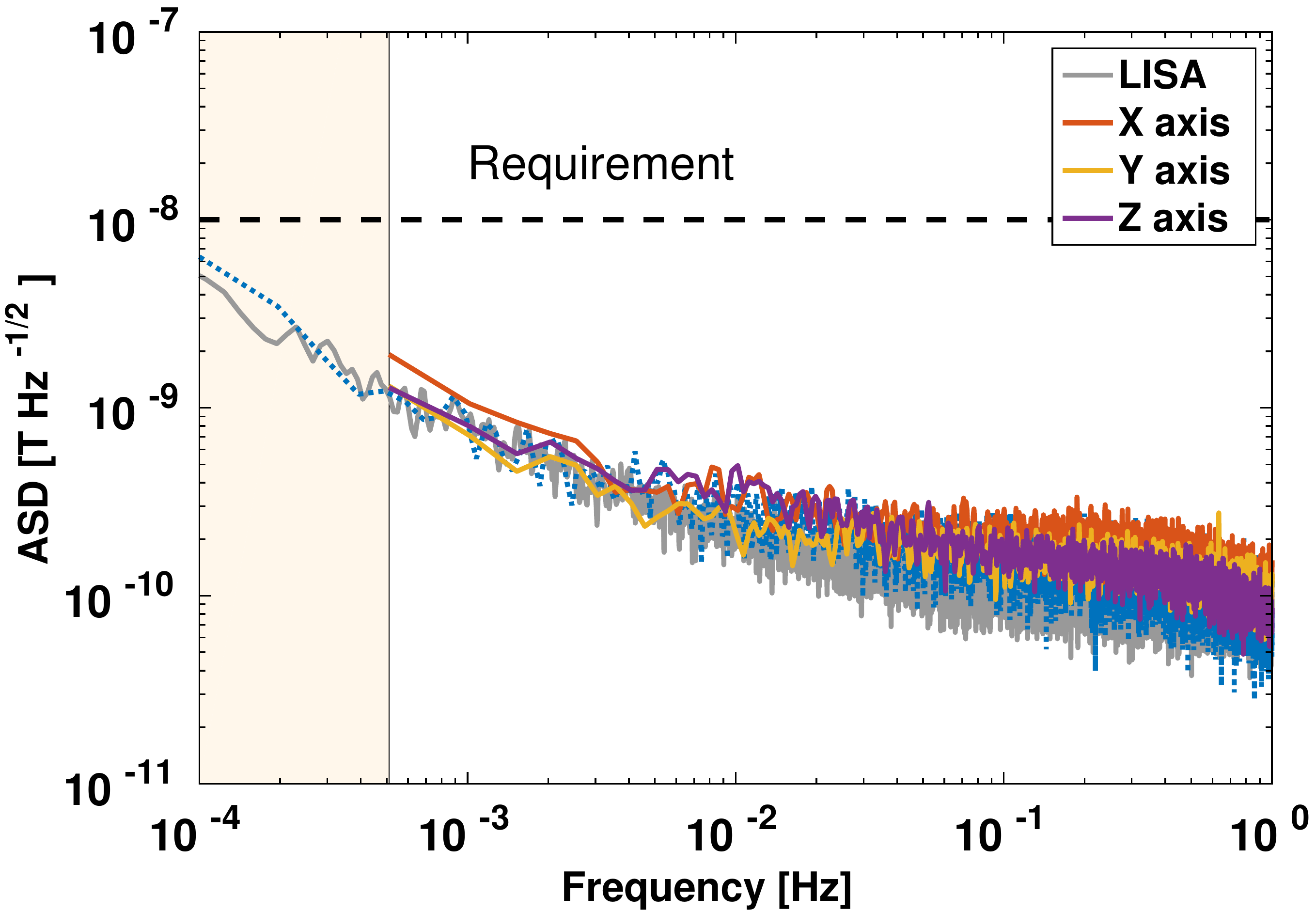}
 \caption{Equivalent magnetic field spectral densities of the CubeSat payload for the 3 h on-ground measurements compared with the current LISA prototype (gray trace\,\cite{bib:AMRpaper}). The dashed blue trace shows the amplitude spectral density of the CubeSat payload extended to lower frequencies (cream-colored area), which can not be assessed in-flight due to CubeSat time constraints. (For interpretation of the references to color in this figure legend, the reader is referred to the web version of this article.)}
 \label{fig:ASDAMRCubeSat}
\end{figure}

On-ground low-frequency noise measurements were taken with the small built-in permalloy magnetic shield. As mentioned before, during in-flight mission operations the duration of the runs for the magnetic experiment  will be no longer than three hours. Therefore, the noise spectral density will be barely estimated down to 1\,mHz. For the lab test in Figure\,\ref{fig:ASDAMRCubeSat}, the noise data of the three sensor axes were taken with a short duration of three hours for comparison purposes during the established mission procedures. As can also be seen in the figure, the equivalent magnetic field noise  is similar to that obtained for the LISA prototype along the CubeSat's operation bandwidth and also when longer noise measurements were extended down to 0.1 mHz. As stated before, the dominant contribution is coming from the intrinsic noise of the sensor itself. More details about the noise performance of the proposed magnetic measurement system for LISA are described in \cite{bib:AMRpaper}.

\begin{figure}[ht!]
  \centering
 \includegraphics[width=0.48\textwidth]{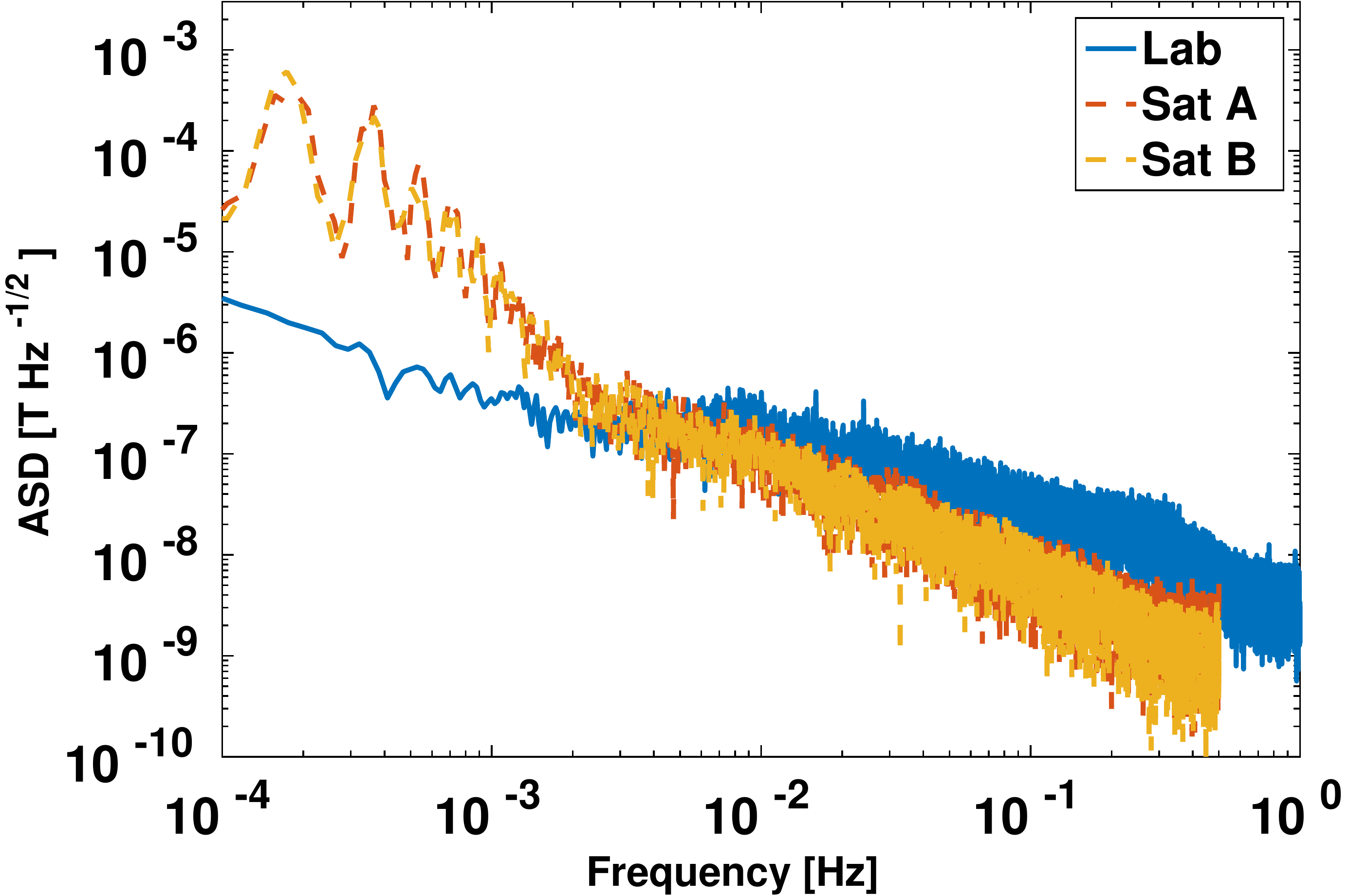}
 \caption{Spectra of the laboratory magnetic environment (solid line) and Swarm vector field magnetometers at two different LEO altitudes (dashed lines). The equivalent noise bandwidth (ENBW) of the spectral window is  $26\,{\rm \upmu Hz}$. (For interpretation of the references to color in this figure legend, the reader is referred to the web version of this article.)}
 \label{fig:LEOvsLab}
\end{figure}

The on-ground noise characterization of the payload was performed under non-controlled external magnetic and thermal environmental conditions. Thence, in order to assess the ability of the in-flight characterization, environmental fluctuations in the laboratory were compared to LEO environmental data obtained with the ESA Swarm satellites at an altitude of $\simeq 445\,{\rm km}$ for Swarm Alpha (Sat A) and $\simeq 510\,{\rm km}$ for Swarm Bravo (Sat B)\,\cite{bib:swarm}. As shown in Figure\,\ref{fig:LEOvsLab}, magnetic conditions in the laboratory are similar or even noisier than the one measured with the Swarm magnetometers above 2 mHz. However, the magnetic noise at lower frequencies for the Swarm data exhibits a contribution that exceeds the environmental noise in the laboratory. These noise contributions are linked to the Earth's magnetic field variations along the spacecraft orbit, which is monitored by the attitude and determination control subsystem (ADCS). Although the geomagnetic field and thermal fluctuations\footnote{Temperature fluctuations in Swarm's vector field magnetometers (Sat A and Sat B) are better than $55\,{\rm mK\,Hz^{-1/2}}$ at $1\,{\rm mHz}$.} are not expected to contribute significantly, further post-processing analyses can be done  during in-flight operation of the CubeSat in order to disentangle their effects.  As a result, we consider the unit together with the magnetic shield capable of performing the noise characterization experiment on-board the CubeSat platform.


%

\subsection{Data storage}

The payload employs two serial EEPROM devices summing a total of 256\,kB of storage capability, where the synchronous serial communication between the uC and the memories is operated via the SPI bus. Once the measurement has been finished, the payload goes into the idle state, switching off the analog signal conditioning circuits (low-power consumption). Then, after receiving the OBC request, the data  are sent from the memories to the OBC through the uC using I$^2$C interface . The data rate of the magnetic measurements from the uC to the EEPROM is 

\begin{eqnarray}
 bitrate &=& \overbrace{4\,{\rm byte}}^{\rm
magnetic \ data}\cdot 8 \cdot \overbrace{5.5}^{f_{\rm s}} = 176 \,{\rm bps},
\end{eqnarray}

\noindent which results in a maximum storage time of 3 hours and 18 minutes per run. 4 byte magnetic data is required because each measurement is the sum of 6144 samples of 16 bits. Regarding the modulation frequency, the value of 5.5 Hz was selected as a trade-off between noise reduction, the effects on additional environmental disturbances produced by a faster switching signal, and the transient response caused by the modulation process. It is important to point out that the noise floor might be reduced to the Johnson noise level of the magnetoresistances by increasing the modulation frequency. However, our interest focused on the low-frequency band and the mitigation of environmental disturbances leads to the selected frequency.

\begin{figure}[ht!]
  \centering
 \includegraphics[width=0.45\textwidth]{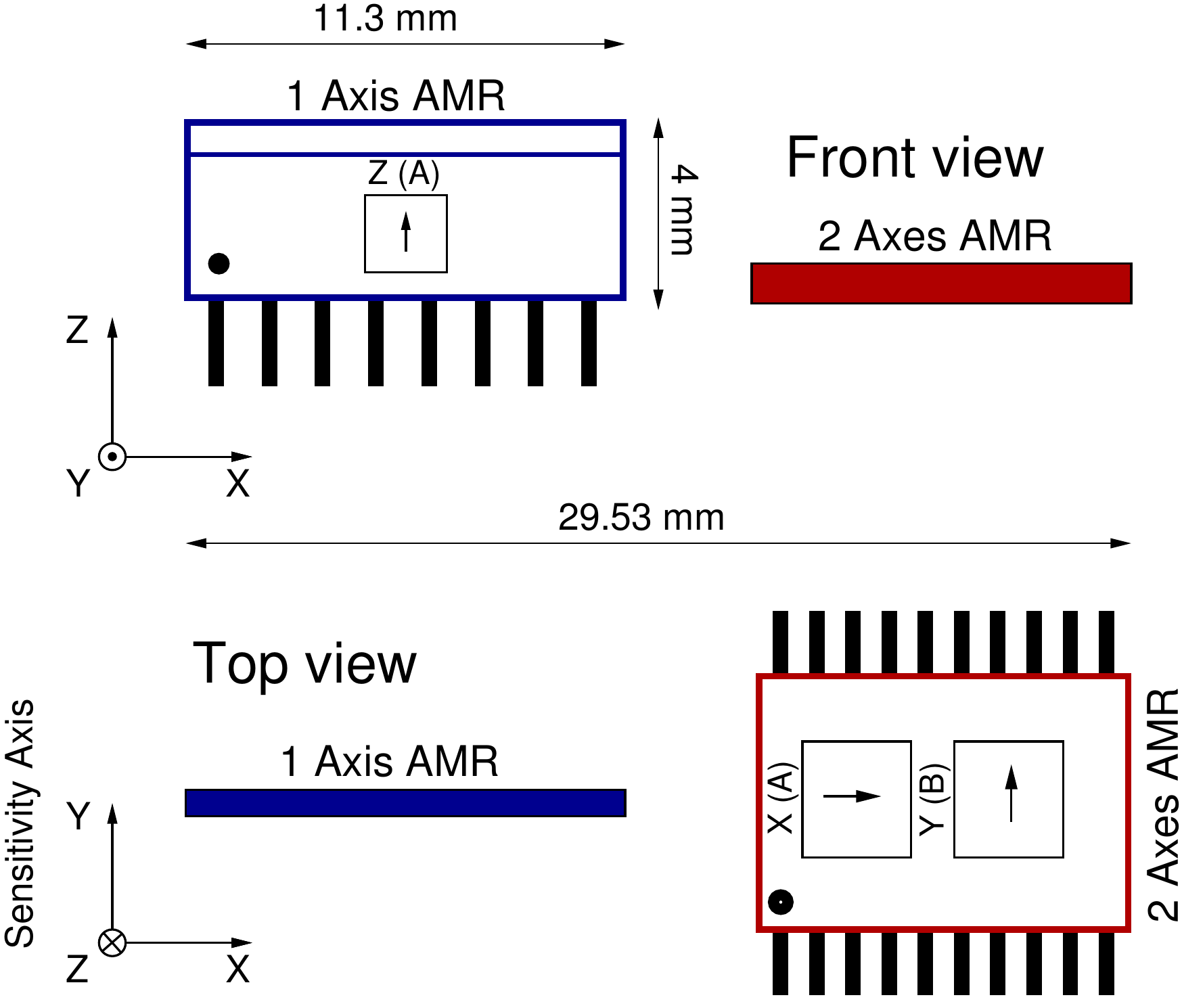}
 \caption [Spatial distribution of the uniaxial and biaxial AMR magnetometers in the CubeSat]{Spatial distribution of the uniaxial and biaxial AMR magnetometers in the CubeSat. Magnetic sensitive axes are displayed. Front (top) and top (bottom) view.}
 \label{fig:AMRlocation}
\end{figure}

\subsection{Sensor head} 

The sensor head is formed by a uniaxial (HMC1001) and a biaxial AMR (HMC1002)\,\cite{bib:honeywell}. The configuration of the magnetometers was mounted directly on the electronics board together with the analog and digital circuits in order to make the integration easier. Conversely, AMR sensors in the current LISA prototype are separated from the electronic readout with the aim of reducing spurious effects. The spatial distribution of the sensor head for the triaxial magnetometer is shown in Figure\,\ref{fig:AMRlocation}, where the two devices were placed along their longitudinal axes so as to reduce the diameter of the innermost shield, which is set by the gap between the height of the uniaxial sensor and the inner layer of the magnetic enclosure.


\subsection{Overall specifications}

\begin{figure}[ht!]
  \centering
 \includegraphics[width=0.47\textwidth]{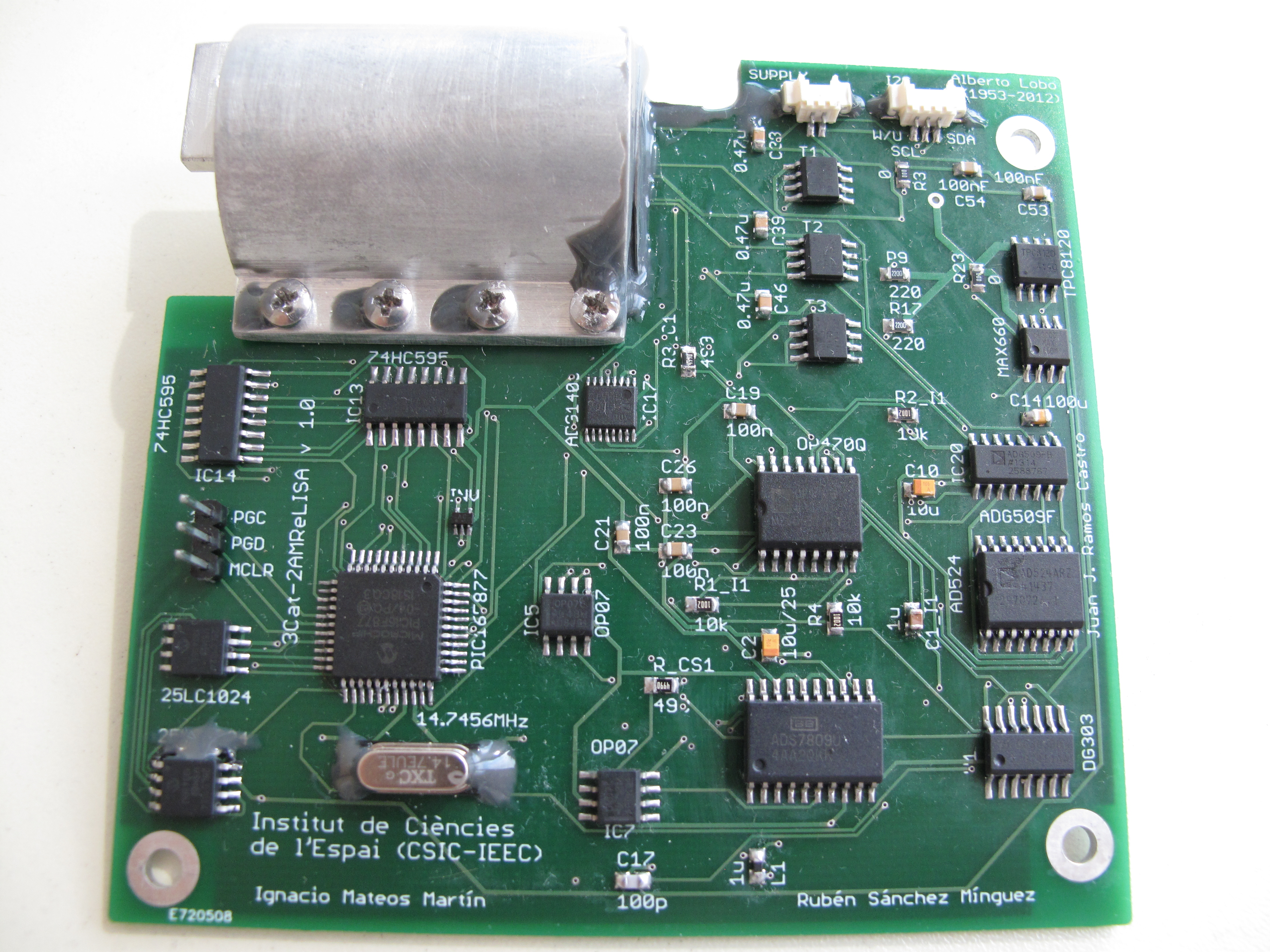}\\
 \includegraphics[width=0.47\textwidth]{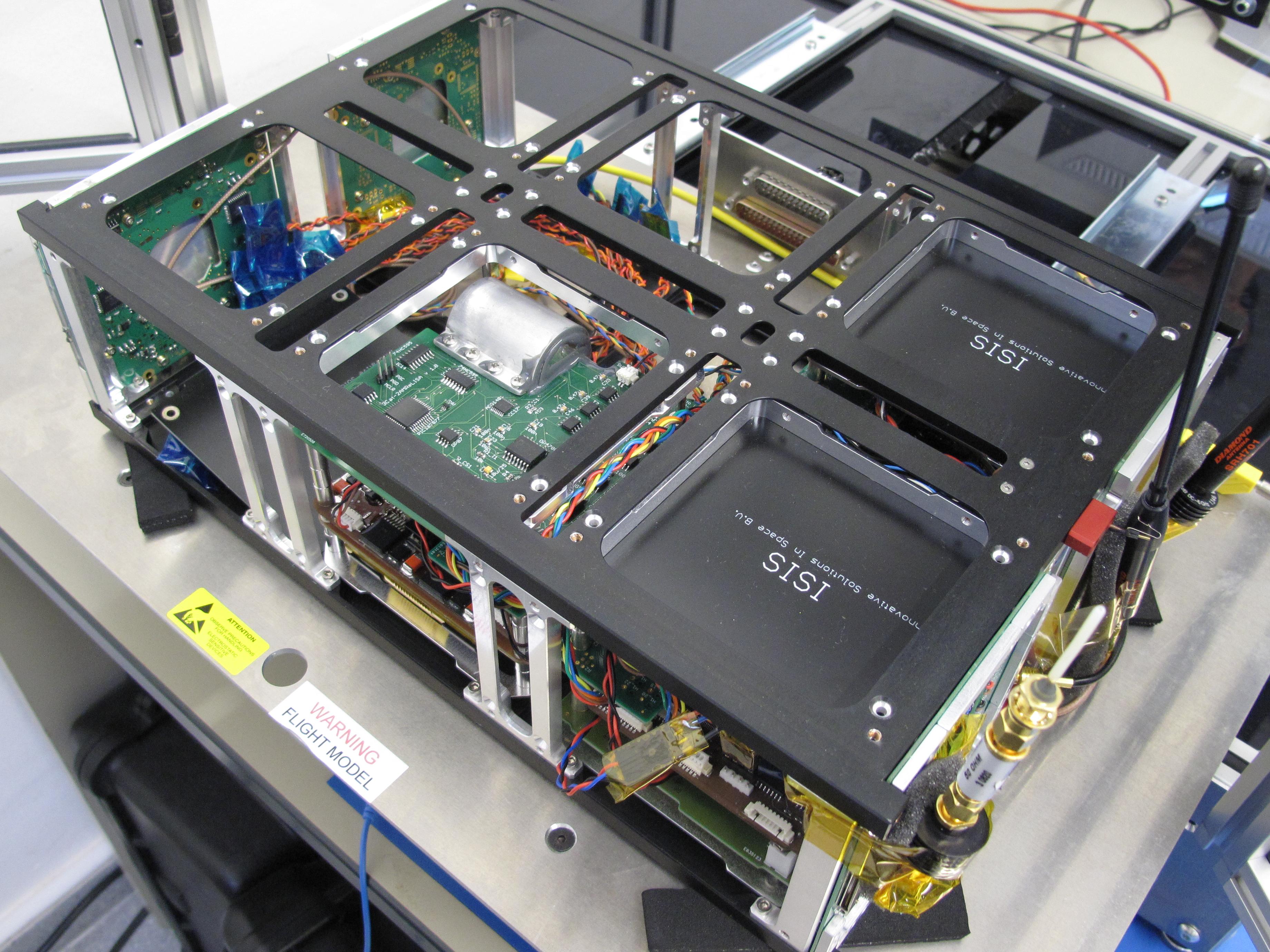}
 \caption {Top: Flight model of the CubeSat payload for the $^3$Cat-2 mission. Bottom: payload integration of the flight model in the CubeSat.}
 \label{fig:FMPayloadCubeSat}
\end{figure}

The overall characteristics of the system were assessed and updated during the three development phases, i.e, the prototype, the engineering qualification model and the final flight model. The instrument meets the noise performance comfortably and the CubeSat requirements specified in Table\,\ref{tab:CubeSatReq}. The flight model of the board with the magnetic shield and its integration in the CubeSat is shown in Figure\,\ref{fig:FMPayloadCubeSat}. 


A general outlook of the specifications for the CubeSat payload is given in Table\,\ref{tab:specCubeSat}. 


\begin{table}[th!]
\caption[Specifications of the magnetic monitoring system for the CubeSat]{Specifications of the magnetic monitoring system for the CubeSat. \label{tab:specCubeSat}}
\centering
\begin{tabular}{l l c}
\hline\hline
\textbf{Parameter} & \textbf{Symb.}  & \textbf{Value} \\
\hline
Field range  & $B_{\rm range}$ & $\pm15\,{\rm \upmu T}$$^{\rm (a)}$  \\
RTI temp. coeff.   & \multirow{2}{*}{$TC_{\rm rti}$} & \multirow{2}{*}{$3\cdot10^{-6}\Delta R_{\rm b,n}\,{\rm V/K}$}   \\
(Sensor + FEE)   & &      \\
\multirow{2}{*}{Noise density}  & \multirow{2}{*}{$S_{\rm B}^{1/2}$} & $0.14\,{\rm  nT}\,{\rm Hz}^{-1/2}$  at 1 Hz  \\
&  & $1\,{\rm  nT}\,{\rm Hz}^{-1/2}$  at 1 mHz \\
Input current & $I_{\rm bridge}$  &  $5\,{\rm mA}$  \\
Input voltage & $V_{\rm bridge}$ & $4.25\, {\rm V}$$^{\rm (b)}$ \\

Linearity error &  &  0.1\% FS  \\

Sensitivity & $s_{\rm AMR}$ & $254.5\,{\rm  mV/\upmu T}$$^{\rm (c)}$\\
ADC resolution & $\Delta B_{\rm ADC}$ &  $0.6\,{\rm nT}$  \\
Equivalent resol. & $\Delta B_{\rm eq.}$ & $0.01\,{\rm nT}$$^{\rm (d)}$\\
Bandwidth & BW & 2.75 Hz  \\
AMR operat. temp. &  & -55$^{\rm o}$C to +150$^{\rm o}$C  \\
Spatial resolution  &   & $< 1$ mm  \\
\multirow{3}{*}{Power consumption}  & $P_{\rm max}$  & $0.32\,{\rm W}$  \\
    &  $P_{\rm nominal}$ & $0.26\,{\rm W}$  \\
 & $P_{\rm standby}$  & $0.025\,{\rm W}$   \\
 Weight  &   & 76.6 g  \\
\hline
\hline
\end{tabular}
\vspace{-0.25cm}
\hspace{2cm}
\begin{flushleft}
 \qquad $^{(a)}${\scriptsize Range limited by the output current of the op-amp.}\\
 \qquad $^{(b)}${\scriptsize For a typical bridge resistance of $850\,\Omega$.}\\
 \qquad $^{(c)}${\scriptsize Typical value for electro-magnetic feedback with a feedback resistor of $499\,\Omega$.}\\
 \qquad $^{(d)}${\scriptsize Oversampling with averaging leads to an equivalent resolution of $16 + \frac{1}{2}\log_2N$ for  $N =  3072$ samples averaged along  each flipping pulse\,\cite{Jespers}. Resolution is limited by AMR noise density.}\\
\end{flushleft}
\end{table}


\section{Conclusions}\label{sec:conclusions}

We have presented an adapted version for CubeSats of the current magnetic measurement system suggested for LISA, the space-based gravitational wave observatory selected as a ESA large-class mission. Chip-scale magnetic sensors emerge as an alternative to the drawbacks found in LISA Pathfinder's magnetometers. In this regard, the main advantages of magnetoresistive sensors are the small size and low-magnetic impact on the surroundings, which allow not only to install more sensors but also to place them closer to the region of interest. The main modifications with respect to the current LISA prototype were carried out in order to fulfill the strict CubeSat constraints regarding power consumption. These changes have effects on some characteristics of the system, such as the magnetic field range, measurement bandwidth, and the simultaneity of the sensors' measurements. However,  they do  not influence the noise performance, where the instrument is comfortably compliant with the envisaged requirement in the measurement bandwidth, which is the most important aspect of the experiment. The technology demonstrator of the LISA magnetic measurement system presented here was successfully integrated in the $^3$Cat-2 CubeSat so as to characterize the low-frequency noise behavior under space environment and elevate the technological maturity of the instrument from the current TRL 4 to the targeted TRL 6. The CubeSat was launched in August 2016 from a Long March 2D Chinese rocket into a Sun-synchronous orbit of around 510-km altitude. Unfortunately, reliable communications between the CubeSat and the ground station could not be established due to a critical failure external to our payload, specifically, a software bug in the OBC. Nevertheless, since the payload was designed in accordance with the CubeSat standard, it can be easily  integrated in other CubeSat platforms employed for testing emerging technologies in space. Currently, its implementation  on the CubeSat developed by the IACTec program at the Institute of Astrophysics of the Canary Islands is under study.

\section*{Acknowledgments}
We wish to thank the UPC NanoSat Lab at ETSETB for the integration of our payload in the $^{3}$Cat-2 CubeSat, the ESA Swarm project for the provided in-flight data of the mission, and Prof. F. de As\'is Mateos and Dr. A. Morales-Garoffolo for the many stimulating and illuminating discussions. Support for the CubeSat payload came from Project AYA2010-15709 of the Spanish Ministry of Science and Innovation (MICINN), ESP2013-47637-P of the Spanish Ministry of Economy and Competitiveness (MINECO), and 2009-SGR-935 (AGAUR).



\end{document}